\documentclass[aps,showpacs,showkeys]{revtex4}
\usepackage{graphicx}
\usepackage{epsfig}
\usepackage{psfrag}
\usepackage{float}
\begin{document}

\title{\bf  Calculated  WIMP signals at the ANDES laboratory: comparison with northern and southern located dark matter detectors. }

\author{O. Civitarese$^1$\footnote{Corresponding author \\ e-mail: osvaldo.civitarese@fisica.unlp.edu.ar}, K. J. Fushimi$^2$ and M. E. Mosquera$^{1,\,2}$}
\affiliation{$^1$\small\it Dept. of Physics, University of La Plata, c.c.~67 1900, La Plata, Argentina.}
\affiliation{$^2$\small\it Facultad de Ciencias Astron\'omicas y Geof\'{\i}sicas, University of La Plata. Paseo del Bosque S/N 1900, La Plata, Argentina.}


\begin{abstract}
Weakly Interacting Massive Particles (WIMP) are possible components of the Universe's Dark Matter. The detection of WIMP is signalled by the recoil of the atomic nuclei which form a detector. CoGeNT at the Soudan Underground Laboratory (SUL) and DAMA at 
the Laboratori Nazionali del Gran Sasso (LNGS)
have reported data on annual modulation of signals attributed to WIMP. Both experiments are located in laboratories of the northern hemisphere. Dark matter detectors are planned to operate (or already operate) in laboratories of the southern hemisphere, like SABRE at Stawell Underground Physics Laboratory (SUPL) in Australia, and DM-ICE in the South Pole.
In this work we have analysed the dependence of diurnal and annual modulation of signals, pertaining to the detection of WIMP, on the coordinates of the laboratory, for experiments which may be performed in the planned new underground facility ANDES
(Agua Negra Deep Experimental Site), to be built in San Juan, Argentina. We made predictions for NaI and Ge-type detectors placed in ANDES, to compare with DAMA, CoGeNT, SABRE and DM-ICE arrays, and found that the diurnal modulation of the signals, at the site of ANDES, is amplified at its maximum value, both for NaI (Ge)-type detectors, while the annual modulation remains unaffected by the change in coordinates from north to south.  
\end{abstract}

\maketitle
\section{Introduction}
\label{Intro}

The observations of Zwicky \cite{zwicky33} and Rubin and Ford \cite{rubin70} demonstrate the existence of dark matter, and therefore have set the motion for an intensive theoretical and experimental search to determine the nature and composition of this new kind of matter. The detection of dark matter is crucial to solve some of the actual cosmological problems. Our present knowledge of the Universe supports the notion that it is composed by ordinary matter, dark matter (DM) and dark energy (DE). The amount of barionic density matter can be determined with great accuracy $\left(5\%\right)$ \cite{planck15,Wmap,Wmap9yr}, and the fractions of the other components amount to $23\%$ (DM) and $72\%$ (DE). An electrically neutral WIMP is the most probably candidate for DM. The estimates of the mass of the WIMP vary from $1 \, {\rm GeV}$ to $10 \, {\rm TeV}$ \cite{freese12}, though this range may be reduced to values of the order of $3 \, {\rm GeV}$ to $20 \, {\rm GeV}$ \cite{davis15}, it interacts weakly and gravitationally with the ordinary matter but does not interact electromagnetically and/or strongly with other particles. 

It is assumed that the DM in the Galactic Halo is composed mostly by WIMP with velocities which obey Maxwell-Boltzmann distribution function \cite{jungman15}. As a consequence of the Earth motion around the Sun, a time variable flux of dark matter (annual modulation) might be detected and from it some properties of the WIMP may be experimentally determined.

There are several efforts to measure the annual modulation of the WIMP's signal, that is experiments which aim at its direct detection \cite{goodman84,wasserman86} like DAMA \cite{bernabei08,bernabei10}, CRESST \cite{angloher12}, CoGeNT \cite{aalseth13,aalseth11}, CDMS \cite{akeriv05, ahmed11}, XENON \cite{aprile11}, SABRE\cite{sabre}, DM-ICE \cite{dmice},
and experiments based on indirect detection methods\cite{silk85,gaisser86} like IceCube \cite{achterberg06}, ANTARES \cite{ageron11}, Fermi-LAT \cite{atwood09}, HESS \cite{aharonian06}, among others.

Because of the extremely low signal-to-noise rates, direct-detection experiments need to be performed in low-background conditions. Therefore, they are located in underground laboratories, and these experiments measure the recoil-energy deposited by WIMP interacting with the atomic nuclei of a detector. To confirm a positive signal, it is important to collect data with the same kind of detectors in two or more different laboratories since, as we shall show, the diurnal modulation of the amplitude of the signal depends on the location of the laboratory. 

Although diurnal modulation of dark-matter signals may be very difficult to measure, if it is observed the difference between data collected in northern and southern locations could help to refine the parameters used to characterize the dark matter. More importantly, a directional detector showing a different declination angle to the WIMP-wind between northern and southern locations  would probe the signal to be caused by dark matter \footnote{We are grateful to the Referee for calling the attention to the relevance of this point}. 

In this work we compare the response of two existing detectors, a NaI based detector (DAMA) and a Ge-type detector (CoGeNT), both located in the northern hemisphere, with the response of similar detectors ideally located in the ANDES laboratory, which will be built in the province of San Juan, Argentina, under the mountains at a deep of $1700\, {\rm m}$ (about 5000 m.w.e) \cite{andesweb,civitarese15,bertou12}. Then, we focus on a similar comparison between NaI-based detectors: (SABRE)\cite{sabre} to operate in a southern location (SUPL, Australia) and DM-ICE\cite{dmice} in the South Pole.

The paper is organized as follows. In Section \ref{formalismo} we describe the formalism needed to compute annual and diurnal modulation rates. In Section \ref{resultados} we demonstrate the dependence of the diurnal modulation with the location of the detector, and make the comparison between ANDES and the other places. Our conclusions are drawn in Section \ref{conclusion}.

\section{Formalism}
\label{formalismo}

The interactions between WIMP and a nucleus of the detector cause the recoil of the nucleus, at the rate (recoil-rate)
\cite{freese12}
\begin{eqnarray}
\label{rate}
\frac{dR}{dE_{\rm nr}}&=&\frac{2\rho_{\chi}}{m_{\chi}}\int{d^3v v f(\vec{v},t) \frac{d\sigma}{dq^2}(q^2,v)}\, ,
\end{eqnarray}
where $E_{\rm nr}= \frac{\mu^2 v^2}{M}(1-\cos\theta)$ is the recoil energy, $\mu=\frac{M m_{\chi}}{M+m_{\chi}}$ is the reduced mass of the system, $M$ is the mass of the nucleus, $m_{\chi}$ is the mass of the WIMP, $\theta$ is the dispersion angle, $\rho_{\chi}=0.3 \, \rm{GeV \, cm}^{-3}$ is the local mass-density of dark matter, $f(\vec{v},t)$ is the WIMP velocity-distribution, and $v$ is the WIMP velocity with respect to the detector. The cross section is written 
\begin{eqnarray}
\label{seccioneficaz}
\frac{d\sigma}{dq^2}(q^2,v)=\frac{\sigma_0}{4 \mu^2 v^2} F^2(q)\, .
\end{eqnarray}
In the previous equation, $q=\sqrt{2 M E_{\rm nr}}$ is the momentum transferred to the nucleus, $\sigma_0$ is the cross section at $q=0$, and $F(q)$ stands for the nuclear form factor.

By defining the mean inverse-velocity
\begin{eqnarray}
\label{inv-vel}
\eta&=&\int{\frac{f(\vec{v},t)}{v}d^3v} \, ,
\end{eqnarray}
the recoil rate of Eq. (\ref{rate}) reads 
\begin{eqnarray}
\label{rate2}
\frac{dR}{dE_{\rm nr}}&=&\frac{2\rho_{\chi}}{m_{\chi}} \frac{\sigma_0}{4 \mu^2 } F^2(q) \eta\, .
\end{eqnarray}

\subsection{Halo component}

In order to obtain the velocity distribution of the WIMP, we have assumed that the model for the DM halo is the Standard Halo Model \cite{freese88}, and, therefore, the velocity distribution is calculated from the truncated Maxwell-Boltzmann distribution \cite{freese12,jungman95}
\begin{eqnarray}
f({\vec{v}})&=&\left\{
\begin{array}{cc}
\frac{1}{N(\pi v_0^2)^{\frac{3}{2}}}e^{{-\left|{\vec{v}}\right|^2}/{v_0^2}} & |\vec{v}|<v_{\rm esc} \\
0&|\vec{v}|>v_{\rm esc} \\
\end{array}
\right. \, , 
\end{eqnarray}
where $N$ is a normalization factor given by
\begin{eqnarray}
N= {\rm erf}\left[\frac{v_{\rm esc}}{v_0}\right]-\frac{2}{\sqrt{\pi}}\frac{v_{\rm esc}}{v_0}e^{-\left(\frac{v_{\rm esc}}{v_0}\right)^2},
\end{eqnarray}
${\rm erf}[x]$ is the error function, $v_{\rm esc}$ and $v_0$ are the escape velocity and the velocity of the Sun, respectively, and their values are $v_{\rm esc}=544 \, {\rm km/s}$ \cite{gelmini15} and $v_0=220 \, {\rm km/s}$ \cite{jungman95}. By adding the laboratory-velocity $\left(\vec{v}_{\rm lab}\right)$, the integral of the Eq. (\ref{inv-vel}) becomes
\begin{eqnarray}
\label{eta1-2}
\eta&=&\frac{1}{N(\pi {v_0}^2)^{3/2}}\int{ \frac{e^{{-({\vec{v}}+{\vec{v}_{\rm lab}})^2}/{v_0^2}}}{v}d^3v}\, .
\end{eqnarray}
After integration, the mean inverse-velocity $\eta$ is written
\begin{widetext}
\begin{eqnarray}
\eta&=&\left\{
\begin{array}{cc}
\frac{1}{2 N v_{\rm lab}}\left(\textrm{erf}\left[\frac{v_{\rm lab}+v_{\rm min}}{v_0}\right]-\textrm{erf}\left[\frac{v_{\rm min}-v_{\rm lab}}{v_0}\right]\right) & v_{\rm min}< v_{\rm esc}-v_{\rm lab}\\
\frac{1}{2 N v_{\rm lab}}\left(\textrm{erf}\left[\frac{v_{\rm esc}}{v_0}\right]-\textrm{erf}\left[\frac{v_{\rm min}-v_{\rm lab}}{v_0}\right]\right) & v_{\rm esc}-v_{\rm lab}< v_{\rm min}< v_{\rm esc}+v_{\rm lab} \\
0 & v_{\rm min}> v_{\rm esc}+v_{\rm lab}
\end{array}\right. \, .
\end{eqnarray}
\end{widetext}
In the previous equation, $v_{\rm min}$ is the WIMP minimal velocity needed to produce the recoil of a nucleus, and its value is \cite{green03}
\begin{eqnarray}
\label{v-min}
v_{\rm min}&=&\sqrt{\frac{E_{\rm nr} (m_{\chi}+M)^2}{2M m_{\chi}^2}}\, .
\end{eqnarray}


\subsection{Laboratory velocity $v_{\rm lab}=\left|\vec{v}_{\rm lab}\right|$}

Since the laboratory, that is the detector, is located at a given position $(\phi_0,\lambda_0)$ on Earth, being $(\phi_0,\lambda_0)$ the latitude and longitude of the site, respectively, we should determine its velocity considering the motion of the Earth and the motion of the Sun, thus, we write 
\begin{eqnarray}
\vec{v}_{\rm lab}&=&\vec{v}_{\odot}^G+\vec{v}_{\oplus}(t,t')\, , \nonumber\\
\vec{v}_{\odot}^G&=&\vec{v}_{\odot}+\vec{v}_{\rm{LSR}}\, , \nonumber\\
\vec{v}_{\oplus}(t,t')&=&\vec{v}_{\rm rev}(t)+ \vec{v}_{\rm rot}(t')\, ,
\end{eqnarray}
where $\vec{v}_{\odot}^G$ is the Sun's velocity with respect to the Galactic System, which is the system of coordinates where the x-axis is pointing to the Galactic Center and the z-axis is pointing to the Galactic North Pole. It can be written as the sum of the peculiar velocity of the Sun with respect to the Local Standard of Rest (LSR), $\vec{v}_{\odot}^G=\left(9,12,7\right) \, {\rm{km/s}}$ \cite{bernabei14}, and the LSR's velocity $\vec{v}_{\rm{LSR}}=\left(0,220 \pm 50,0\right) \, {\rm{km/s}}$ \cite{bernabei14}. Both velocities are defined using vector-components relative to the Galactic System of coordinates. The term $\vec{v}_{\oplus}(t,t')$ is the sum of the orbital velocity of the Earth $\vec{v}_{\rm rev}(t)$ and the rotational velocity $\vec{v}_{\rm rot}(t')$. The time $t$ is expressed in sidereal days, and $t'$ in sidereal hours.

\begin{figure}[H]
\begin{center}
\epsfig{file=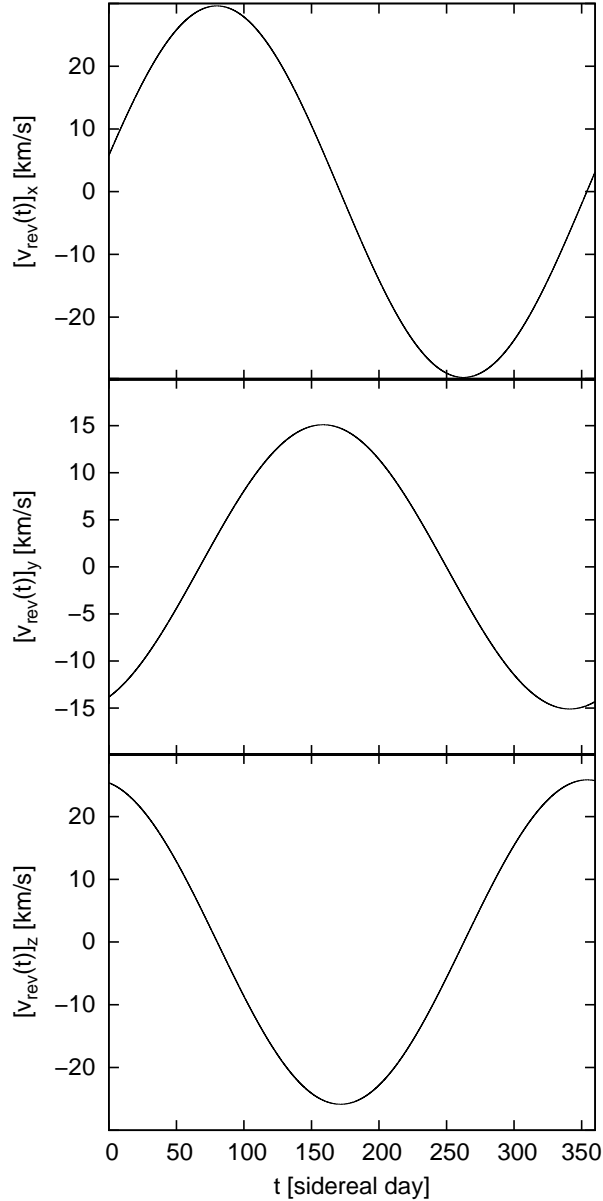, angle=-90, width=240pt}
\end{center}
\caption{Components of the orbital-velocity as a function of the sidereal time.} \label{comp-vrev}
\end{figure}

For the mean orbital-velocity we adopt the value: $v_{\rm rev}^{\oplus}= 29.8\, {\rm{km/s}}$ \cite{bernabei14}, and the value of the orbital (annual)-velocity is given by the expression 
\begin{eqnarray}
\vec{v}_{\rm rev}(t)&=& v_{\rm rev}^{\oplus} \left[{\varepsilon}^{\rm eclip}_1 \sin(w_{\rm rev} (t-t_{\rm eq}))
-{\varepsilon}^{\rm eclip}_2 \cos(w_{\rm rev}(t-t_{\rm eq}))\right] \, .
\end{eqnarray}
In the previous equation $w_{\rm rev}$ is the orbital frequency, $t_{\rm eq}$ is the sidereal time of March-equinox and, in the Galactic System,
\begin{eqnarray}
{\varepsilon}^{\rm eclip}_1 &=&(-0.055, 0.494, -0.867)\,,\nonumber\\
{\varepsilon}^{\rm eclip}_2 &=&(-0.993, -0.112, -2.58\times 10^{-4})\,,\nonumber\\
{\varepsilon}^{\rm eclip}_3 &=&(-0.097, 0.862, 0.497)\,.
\end{eqnarray}
In Fig. \ref{comp-vrev} we show the components of the orbital velocity as a function of sidereal time.

\begin{figure}[H]
\begin{center}
\epsfig{file=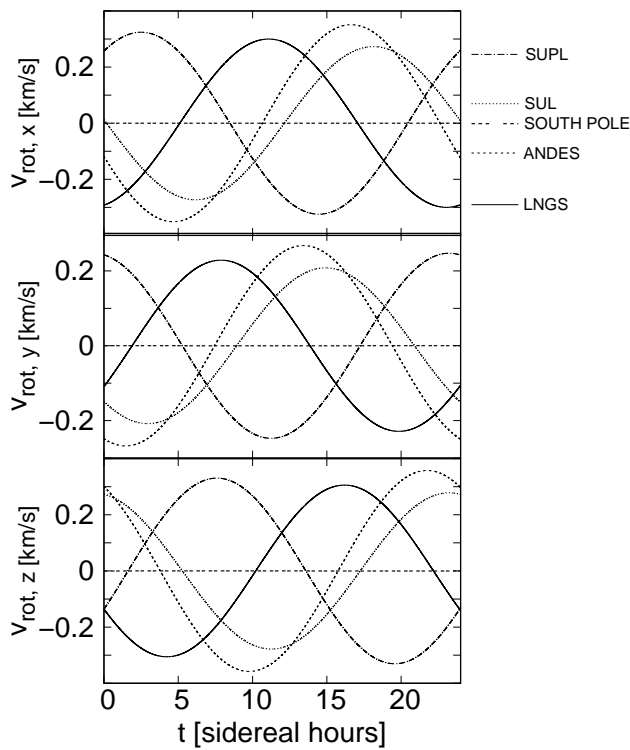, angle=0, width=240pt}
\end{center}
\caption{Rotational velocity of the experimental sites (see the text) as a function of sidereal time.} \label{comp-vrot}
\end{figure}

The Earth rotational (day)-velocity is written as
\begin{eqnarray}
\vec{v}_{\rm rot}(t')&=& v_{\rm rot}^{\oplus} \left[{\varepsilon}^{\rm equ}_1 \sin(w_{\rm rot}(t'+t_{0})-{\varepsilon}^{\rm equ}_2 \cos(w_{\rm rot}(t'+t_{0}))\right] \, ,
\end{eqnarray}
where $v_{\rm rot}^{\oplus}= V_{\rm equ}\cos(\phi_0)$, and $V_{\rm equ}=0.4655 \, {\rm{km/s}}$ is the value of the rotational velocity in the Equator \cite{bernabei14}. The frequency is $w_{\rm rot}$ and $t_{0}$ is the time associated to the laboratory longitude $\lambda_0$. Respect to the Galactic System of reference the vectors ${\varepsilon}^{\rm equ}_k$ are written
\begin{eqnarray}
{\varepsilon}^{\rm equ}_1 &=&(-0.055, 0.494, -0.867)\, ,\nonumber\\
{\varepsilon}^{\rm equ}_2 &=&(-0.873, -0.446, -0.198)\, ,\nonumber\\
{\varepsilon}^{\rm equ}_3 &=&(-0.485, 0.746, 0.456)\, .
\end{eqnarray}
In Fig. \ref{comp-vrot} we show the components of the rotational velocity as a function of sidereal time for labs in the northern and southern hemisphere, that is the LNGS in Italy \cite{lngsweb,bernabei14}, which is the location of DAMA \cite{bernabei14} and SUL  in USA, which is the location of the CoGeNT detector \cite{cogentweb}, in the north, and ANDES \cite{andesweb,civitarese15,bertou12,machado12}, SUPL, which is the place of the SABRE detector in Australia \cite{sabre}, and the South Pole, which is the place of DM-ICE \cite{dmice} in the south, respectively. The coordinates of each site, longitude and latitude, are listed in Table \ref{lab}.

\begin{table}
\begin{tabular}{ccc}
\hline
\textbf{Laboratory} & $\phi_0$ & $\lambda_0$ \\ \hline
LNGS & $42^{\circ}27'$ N & $13^{\circ}34'$ E \\ \hline
SUL & $47^{\circ}48'$ N & $92^{\circ}14'$ W \\ \hline
ANDES & $30^{\circ}15'$ S & $69^{\circ}53'$ W \\\hline
SUPL & $37^{\circ}3'$ S & $142^{\circ}46'$ E \\\hline
South Pole & $89^{\circ}59'$ S & $139^{\circ}16'$ E \\\hline
\end{tabular}
\caption{Latitude ($\phi_0$) and longitude ($\lambda_0$), of the LNGS, SUL, ANDES, SUPL and South Pole places.} \label{lab}
\end{table}

With the previous definitions, of the orbital and rotational velocities, the laboratory velocity can be written as
\begin{eqnarray}
v_{\rm lab} &\simeq& \vert{\tilde{\vec{v}}_{\odot}^G} \vert + v_{\rm rev}^{\oplus} A_{\rm m} \cos(w_{\rm rev}(t-\tilde{t_0})) +v_{\rm rot}^{\oplus} A_{\rm d} \cos(w_{\rm rot}(t'-t_{\rm d}))\, ,
\end{eqnarray}
where $\check{v}_{\rm lab}=\frac{\vec{v}_{\rm lab}}{\vert{\vec{v}_{\rm lab}}\vert}$, $\vert{\tilde{\vec{v}}_{\odot}^G} \vert=\vert{\vec{v}_{\odot}^G} \vert +\frac{1}{2} \frac{\vert\vec{v}_{\rm rev}\vert ^2}{\vert{\vec{v}_{\odot}^G} \vert}$, $\tilde{t_0}= t_{\rm eq}+\frac{\beta_{\rm m}}{w_{\rm rev}}$, $A_{\rm m}=0.488$, $\beta_{\rm m}=1.260\, {\rm rad}$, $t_{\rm d}=\frac{\beta_{\rm d}}{w_{\rm rot}}-t_{0}$, $A_{\rm d}=0.6712$ and $\beta_{\rm d}=3.9070 \, {\rm rad}$ \cite{bernabei14}.

\begin{figure}[H]
\begin{center}
\epsfig{file=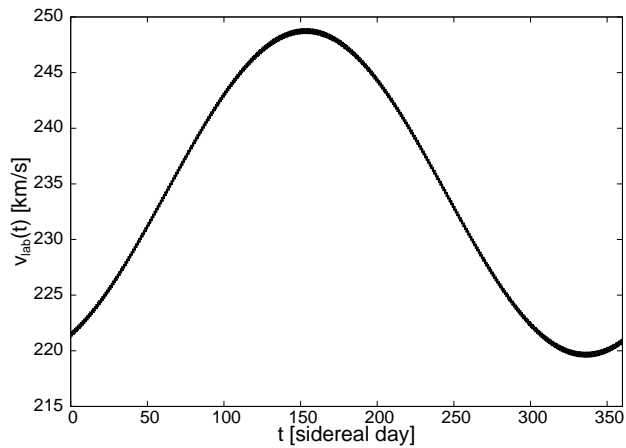, angle=-90, width=240pt}
\end{center}
\caption{Laboratory velocity as function of the sidereal time, for a year-long interval. In this scale the curves corresponding to each lab do not differentiate each other (see the text for the explanation).}
\label{vlab}
\end{figure}

\begin{figure}[H]
\begin{center}
\epsfig{file=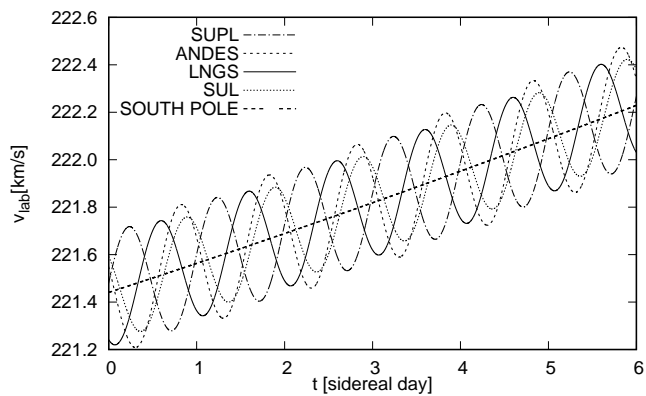, angle=0, width=240pt}
\end{center}
\caption{Laboratory velocity as function of the sidereal time, for a shorter period of time.}
\label{vlab-zoom}
\end{figure}

In Figures \ref{vlab} and \ref{vlab-zoom} we show the laboratory velocity, for each site, as function of the sidereal time. The curves in Fig. \ref{vlab} do coincide, apparently, but if the time-scale and the velocity values are depicted in smaller steps, they differ from each other, as shown in Fig. \ref{vlab-zoom}. 

\subsection{Nuclear form factor}

We have considered the spin-independent cross section
\begin{eqnarray}
\sigma_0^{\rm{SI}}&=&\frac{4}{\pi} \mu^2 \left[\rm{Z }\rm{f}_p+(\rm{A}-\rm{Z})\rm{f}_n\right]^2\, ,
\end{eqnarray}
where $Z$ is the proton number, $A-Z$ the neutron number, $\rm{f}_p$ and $\rm{f}_n$ are the proton and neutron coupling factors. The nuclear form-factor can be defined as \cite{chen11}
\begin{eqnarray}\label{fq}
F(q)&=&\frac{4\pi}{A} \int_0^\infty {\frac{r}{q}\rho(r)\sin(qr)dr} \, ,
\end{eqnarray}
where $\rho(r)$ is the nuclear density
\begin{eqnarray}
\rho(r)&=&\frac{\rho_0}{1+e^{\frac{(r-R_0)}{a}}}\, ,
\end{eqnarray}
and where $a \simeq 0.65\, \rm{fm}$, $R_0=1.12 A^{\frac{1}{3}}\, \rm{fm}$ and $\rho_0=0.17 \,\rm{nucleons} /\rm{fm}^{3}$ \cite{bohr}.

In order to obtain an analytic expression of the form factor we have performed the expansion of $\sin(qr)$, in Eq.(\ref{fq}), leading to
\begin{eqnarray}
\label{factor}
F_{\rm{WS}}(q)&=&\frac{4\pi \rho_0 a}{A} \sum_{k=1}\frac{(-1)^{k+1}q^{2k-2}}{(2k-1)!}\int_{-\frac{R_0}{a}}^\infty{\frac{\left(a z+R_0\right)^{2k}}{1+e^z} dz} \, .\nonumber \\ 
\end{eqnarray}
The integral of the last expression can be divided into two integrals by expanding 
\begin{equation}
(za+R_0)^{2k}=\sum_{j=0}^{2k} \frac{(2k)!}{j!(2k-j)!} a^{2k-j}R_0^j z^{2k-j}
\end{equation}
and writing 
\begin{eqnarray}
\int_{-\frac{R_0}{a}}^\infty{\frac{z^{2k-j}}{1+e^z} dz} &=&\int_{0}^\infty { \frac{z^{2k-j}}{1+e^z} dz}+\int_{\frac{-R_0}{a}}^0 {\frac{z^{2k-j}}{1+e^z} dz}\, .\nonumber
\end{eqnarray}
After these replacements the form factor, Eq.(\ref{factor}), reads
\begin{widetext}
\begin{eqnarray}
F_{\rm{WS}}(q)&=&\frac{4\pi \rho_0 a^3}{A} \sum_{k=1}\frac{(aq)^{2k-2}}{(2k-1)!} \sum_{j=0}^{2k} \frac{(2k)!}{(2k-j)!j!} \left(\frac{R_0}{a}\right)^j \Bigg\{\sum_{n=1}^{\infty}\frac{(-1)^{n+k}}{n^{2k-j+1}}\Gamma(2k-j+1)+(-1)^{j+k+1}\frac{({\frac{R_0}{a}})^{2k-j+1}}{2k-j+1}\nonumber\\
&&\hskip 6cm +\sum_{n=1}^{\infty}\frac{(-1)^{n-j+k+1}}{n^{2k-j+1}}\left[\Gamma(2k-j+1)-\Gamma\left(2k-j+1,\frac{nR_0}{a}\right)\right]\Bigg\} \, . \nonumber\\
&&
\end{eqnarray}
\end{widetext}

The form factor obtained using a Woods-Saxon density is similar to the empirical form factor of Helm \cite{helm56}
\begin{eqnarray}
 F_{\rm H}(q)&=&3 e^{-q^2s^2/2}\left ( \frac{\sin{(qr_n)}-qr_n\cos{(qr_n)}}{(qr_n)^3} \right) \,
\end{eqnarray}
where $s \simeq 0.9\, {\rm fm}$, $r_n^2=c^2+\frac{7}{3}\pi^2a^2-5s^2$, $a \simeq 0.52\, {\rm fm}$ and $c \simeq \left(1.23 A^{1/3}- 0.6 \right)\, {\rm fm}$ \cite{lewin96}.


\subsection{Annual and diurnal modulation rates}

We can write the recoil rate in terms of the annual and diurnal modulation rates as
\begin{eqnarray}
\frac{dR}{dE_{\rm nr}}&=& S_0+S_{\rm m}(E_{\rm nr})\cos(w_{\rm rev}(t-\tilde{t_0})) 
+S_{\rm d}(E_{\rm nr})\cos(w_{\rm rot}(t'-t_{\rm d}))\, , \label{recoil}
\end{eqnarray}
where the annual modulation is given by
\begin{eqnarray}
\label{sm}
S_{\rm m}(E_{\rm nr})&=& \frac{\rho_{\chi}}{m_{\chi}}\frac{\sigma_0}{2\mu^2}F^2(q) v_{\rm rev}^{\oplus}A_{\rm m} \frac{\partial \eta}{\partial v_{\rm lab}}\Bigg|_{\tilde{t_0};t_{\rm d}} \, ,
\end{eqnarray}
and the diurnal modulation is
\begin{eqnarray}
\label{sd}
S_{\rm d}(E_{\rm nr})&=&\frac{\rho_{\chi}}{m_{\chi}}\frac{\sigma_0}{2\mu^2}F^2(q)v_{\rm rot}^{\oplus}A_{\rm d}\frac{\partial \eta}{\partial v_{\rm lab}}\Bigg|_{\tilde{t_0};t_{\rm d}}\, .
\end{eqnarray}

The average modulation amplitudes are defined as
\begin{eqnarray}
< S_{\rm m}>&=&\frac{1}{E_2-E_1}\int_{E_1}^{E_2} S_{\rm m}(E_{\rm nr})dE_{\rm nr} \, ,\\
\label{sdmedia}
< S_{\rm d}>&=&\frac{1}{E_2-E_1}\int_{E_1}^{E_2} S_{\rm d}(E_{\rm nr})dE_{\rm nr}\, ,
\end{eqnarray}
for recoil energies in the domain $(E_1,E_2)$.
\section{Results}
\label{resultados}

We have adopted best-fit values, of the WIMP mass and cross-section, which have been reported by the existing collaborations, in order to compute DM signals measured by a detector located in ANDES laboratory.

As we can see from Eq. (\ref{sd}), the recoil-rate depends on the location of the detector. The phase of the diurnal modulation amplitude, $t_{\rm d}=\frac{\beta_{\rm d}}{w_{\rm rot}}-t_{0}$ of Eq. (\ref{recoil}), depends on the longitude of the experimental site, while the latitude changes the value of the amplitude of the diurnal modulation. Because of this dependence with the coordinates of the laboratory, we present same predictions of the average diurnal modulation amplitude, for NaI-and Ge-type detectors placed in ANDES \cite{civitarese15}, and compare the results with the values corresponding to the labs of the northern and southern 
hemispheres. In all cases we shall present the results for the calculated signals, without making explicit comparisons with  experimental data, because the aim of the calculations is to show the advantage (or disadvantage) of ANDES's locations respect to the other places. 


\subsection{NaI detector}

In order to compute the average diurnal modulation amplitude, we use the parameters obtained in Ref. \cite{freese12,savage08}, that is $m_\chi = 11 \,{\rm GeV}$, and $\sigma_0^{\rm{SI}}=2 \times 10^{-14} \, {\rm fm}^2$, and calculate the recoil of the Na nuclei. 

\begin{figure}[H]
\begin{center}
\epsfig{file=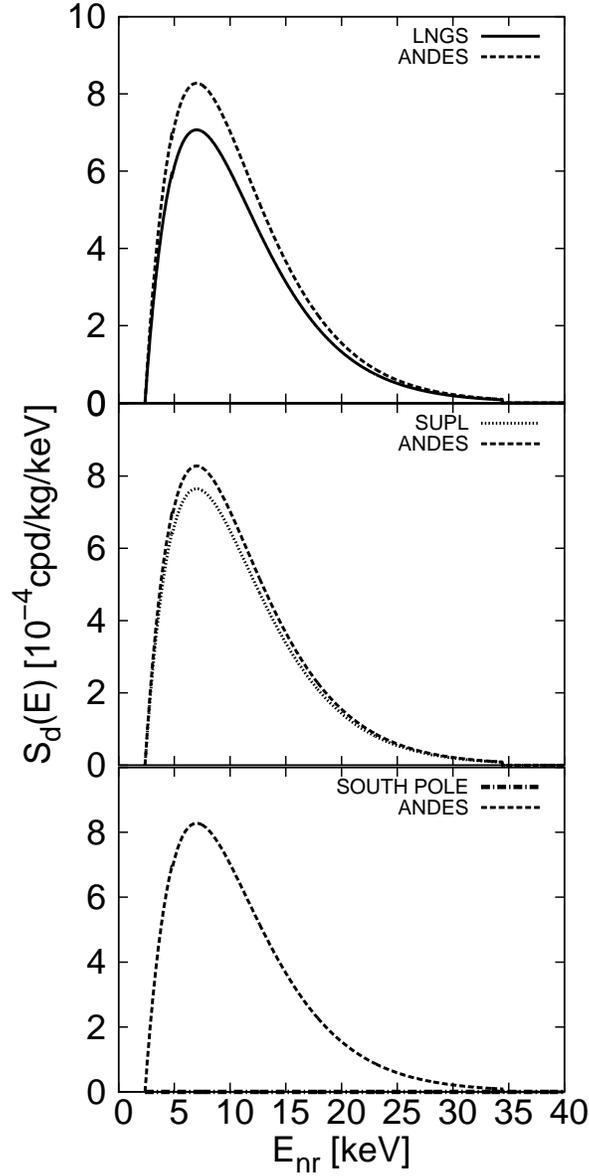, angle=0, width=240pt}
\end{center}
\caption{Diurnal modulation amplitude as a function of the recoil energy, in units of cpd/Kg/keV (counts per days per Kg of material
and per keV). The insets show the calculated values corresponding to ANDES and to the other laboratory-locations, as indicated in the figures. }
\label{modulaciondiurna-vs-energia}
\end{figure}

In Figure \ref{modulaciondiurna-vs-energia} we show the results for the diurnal modulation amplitude, as a function of the nuclear recoil energy, for the interaction of WIMP with the Na nuclei of the detector.
As one can notice, the diurnal modulation amplitude is larger for the detector in ANDES, compared to the calculated signals for the other labs which operate NaI detectors. As mentioned before, the annual modulation amplitude will be the same for experiments performed in different laboratories.

\begin{figure}[H]
\begin{center}
\epsfig{file=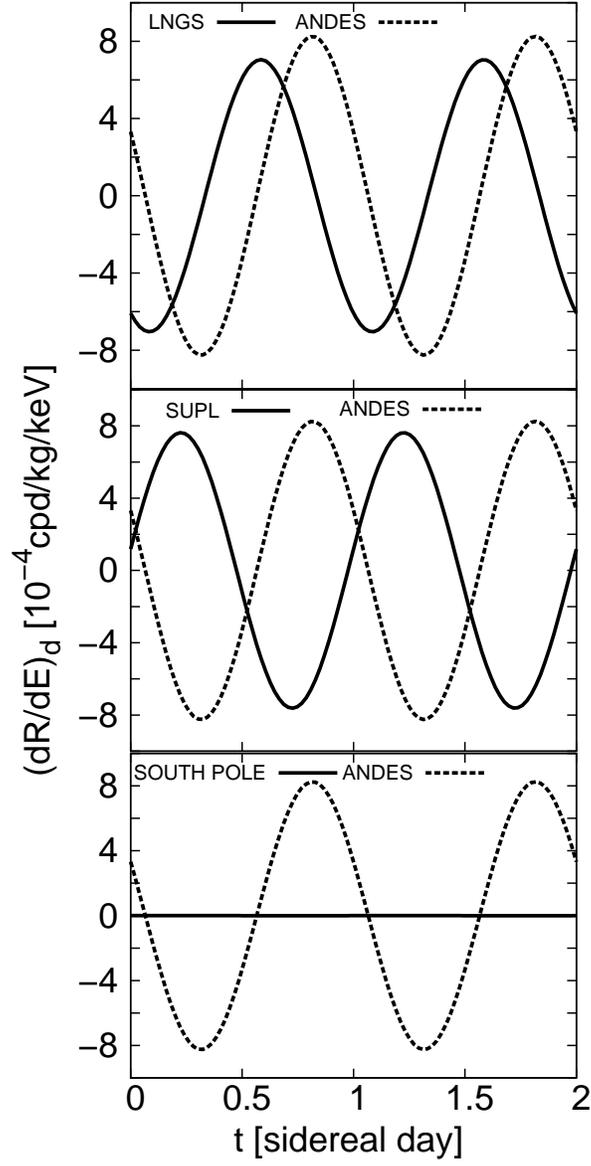, angle=0, width=240pt}
\end{center}
\caption{Calculated diurnal recoil-rate, as a function of sidereal time, for the three sites which operate NaI detectors, and ANDES.}
\label{modulacion-tiempo-dama}
\end{figure}

The comparison of data and theoretical results requires a correction to the recoil energy. The energy that the detector measures, $E$, and the nuclear recoil, $E_{\rm nr}$, are related by a quenching factor, $Q$, such that $E=Q E_{\rm nr}$, where the actual value of the quenching factor depends on the type of detectors. For the NaI-type detector the quenching factor is $Q_{Na}=0.3$ \cite{bernabei98}. In Figure \ref{modulacion-tiempo-dama} we present the diurnal modulation as a function of the sidereal time (in days) for an energy $E=2 \, {\rm keV}$. As one can see, there exists a noticeably shift between the calculated signals. The amplitude of the modulation is larger for the case of the ANDES laboratory, compared with the results corresponding to LNGS and South Pole
detectors, and it is quite comparable for ANDES and SUPL. 

\begin{figure}[H]
\begin{center}
\epsfig{file=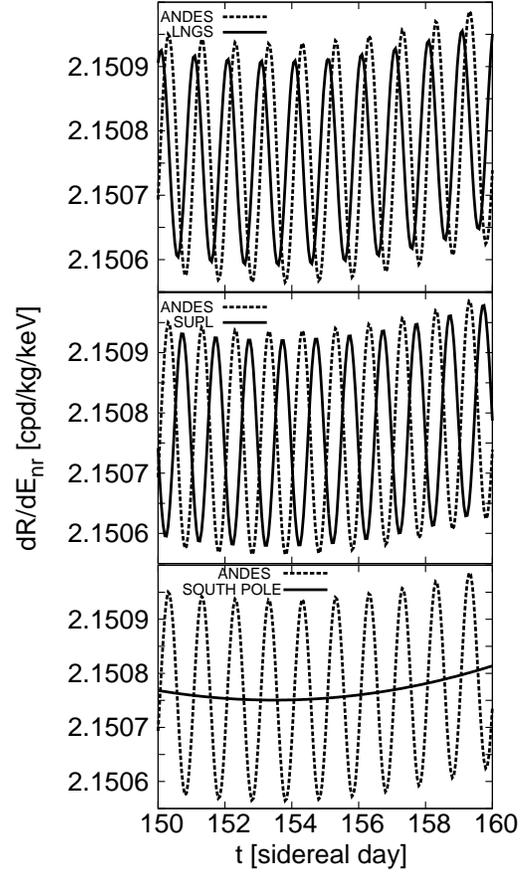, angle=0, width=240pt}
\end{center}
\caption{Total recoil rate, at the minimum, and for a nuclear recoil energy of $E_{nr}=2$ keV.}
\label{xx}
\end{figure}

\begin{figure}[H]
\begin{center}
\epsfig{file=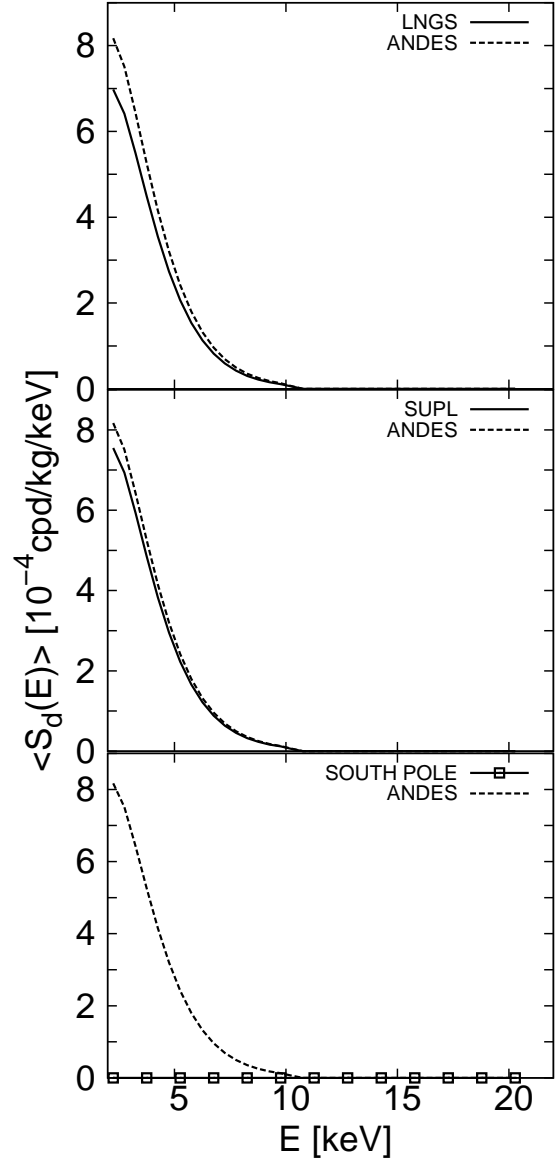, angle=0, width=240pt}
\end{center}
\caption{Average diurnal modulation amplitude, for the recoil of Na nuclei, as a function of the energy.}
\label{modulaciondiurna-DAMA}
\end{figure}

Figure \ref{xx} shows the total recoil-rate, at the minimum, for a longer period of time, and for $E_{nr}=2\, \rm keV$. It is seen that the values corresponding to ANDES are systematically larger than those of the other locations,
and that there is a shift, more pronounced for SUPL, between the calculated signals. In Figure \ref{modulaciondiurna-DAMA} we present results for the average modulation amplitude as a function of the energy. The values corresponding to ANDES are larger than the LNGS and South-Pole ones, and they are of comparable magnitude respect to the Australian site
(SUPL).

To give an idea about the effects on the measurements, that is the difference in counts due to the location of the detector, in Table \ref{tablenai} we show the calculated values for the expected number of counts for NaI detectors, at intervals of 0.5 and 1 sidereal day, respectively. The values given in this table have been extracted from the curves of Figure \ref{modulacion-tiempo-dama}.

\begin{table}[H]
\centering
\caption{NaI detectors}\label{tablenai}
\begin{tabular}{c|c|c|}
\hline
 lab          & 0.5 sidereal day [$10^{-4}$ \rm {cpd/kg/keV}] & 1 sidereal day [$10^{-4}$ cpd/kg/keV] \\
\hline
ANDES      & $-3.32018$                                                    &$ 3.32018$                                                 \\
LNGS       & $6.08538$                                                     & $-6.08538$                                                  \\
SUPL       & $-1.18185$                                                    & $1.18185$                                                   \\
SOUTH POLE & $-0.0000868153$                                               & $0.0000868153 $                                            
\\
\hline
\end{tabular}
\end{table}

\subsection{Ge detector}

\begin{figure}[H]
\begin{center}
\epsfig{file=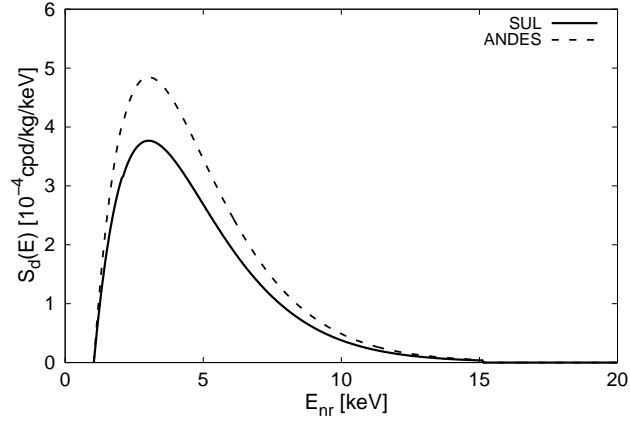, angle=-90, width=240pt}
\end{center}
\caption{Diurnal modulation amplitude (Ge detector) as a function of the recoil energy.}
\label{modulaciondiurna-vs-energia-cogent}
\end{figure}

The energy-dependence of the diurnal modulation-amplitude, for experiments at SUL and ANDES, is shown in Figure \ref{modulaciondiurna-vs-energia-cogent}. The calculations have been performed accounting for the recoil of Ge, and using 
the parameters given in Ref. \cite{aalseth14}, that is $m_\chi = 10 \,{\rm GeV}$, and $\sigma_0^{\rm{SI}}= 10^{-15} \, {\rm fm}^2$. 
Also for this type of detectors, the diurnal modulation amplitude is larger for the detector placed in ANDES.

To determine the average diurnal modulation amplitude we use the quenching factor for Ge, $Q_{Ge}=0.2$ and the response function \cite{savage08}
\begin{eqnarray}
\epsilon \left(E\right) &=& 0.66-\frac{E}{50 \,{\rm keV}} \, ,
\end{eqnarray}
to perform the integral of Eq. (\ref{sdmedia}). 

\begin{figure}[H]
\begin{center}
\epsfig{file=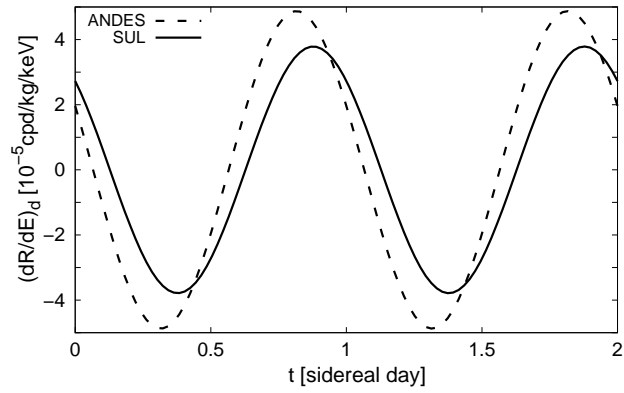, angle=-90, width=240pt}
\end{center}
\caption{Contribution of the diurnal modulation to the recoil rate, as a function of the sidereal time.}
\label{modulacion-tiempo-cogent}
\end{figure}

\begin{figure}[H]
\begin{center}
\epsfig{file=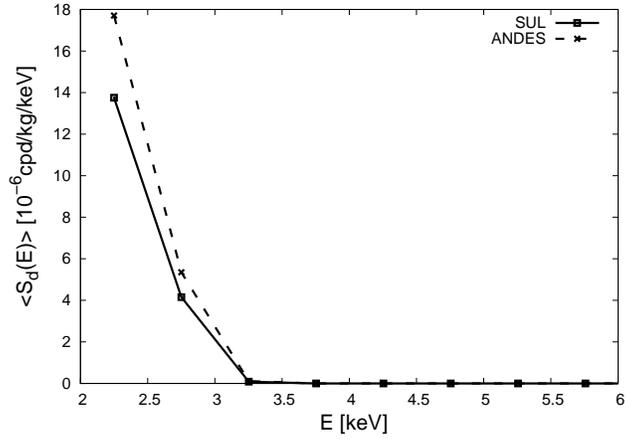, angle=-90, width=240pt}
\end{center}
\caption{Average diurnal modulation on ANDES and SUL experimental sites.}
\label{modulaciondiurna-cogent}
\end{figure}

In Figure \ref{modulacion-tiempo-cogent} we show the diurnal modulation contribution to the recoil rate as a function of the sidereal time (in days) for an energy of $E=2 \, {\rm keV}$. As it is seen from the figure it exists a shift between the signals and the amplitude is larger for the detector placed at ANDES. 
\begin{figure}[H]
\begin{center}
\epsfig{file=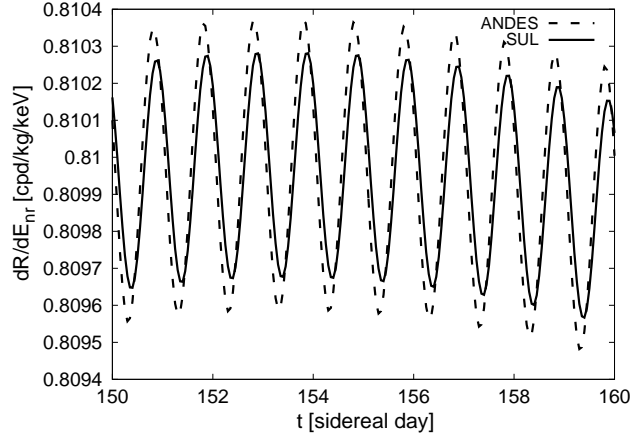, angle=-90, width=240pt}
\end{center}
\caption{Calculated recoil rate, at the maximum, as a function of the sidereal time (Ge detector).}
\label{xxx}
\end{figure}
Figure \ref{xxx} shows the recoil-rate, at the maximum and for an energy $E_{nr}=2 \, {\rm keV}$. In Figure \ref{modulaciondiurna-cogent} we present the results for the SUL and ANDES sites for the average modulation amplitude as a function of the recoil energy. As it is seen from this figure, at the maximum, the results for ANDES are larger by a factor of the order of $1.29$, respect to the results of SUL.

Finally, In Table \ref{tablege} we show the calculated values for the expected number of counts for Ge detectors, at intervals of 0.5 and 1 sidereal day. The values given in this table have been extracted from the curves of Figure \ref{modulacion-tiempo-cogent}.

\begin{table}[H]
\centering
\caption{Ge detectors}\label{tablege}
\begin{tabular}{c|c|c|}
\hline
 lab  & 0.5 sidereal day [$10^{-5}$ cpd/kg/keV] & 1 sidereal day [$10^{-5}$ cpd/kg/keV] \\
\hline
ANDES & $-1.96143 $                                                   & $1.96143$                                                   \\
SUL   & $-2.72758  $                                                  & $2.72758 $     
\\
\hline                                         
\end{tabular}
\end{table}

\section{Conclusions}
\label{conclusion}

In this work we have calculated observables associated to the scattering of WIMPs by nuclei, like amplitudes for diurnal and annual modulations and recoil-rates. We have considered two different detectors, NaI and Ge, and we have shown that, for the best-fit values of the WIMP mass and cross section, the diurnal modulation amplitude depends on the location of the detector on the Earth. We have compared the expected signals, for both type of detectors and taken the coordinates of existing underground facilities, with  those of 
a WIMP detector to be located in the Agua Negra Deep Experimental Site (ANDES underground laboratory). We found that the value of the average diurnal modulation, for NaI and Ge detectors placed in ANDES, will be larger than the values obtained for detectors placed in other labs. The enhancement favouring ANDES location correlates with the rate of the latitude's cosine of the sites.

\section*{{\bf Acknowledgments}}
Support for this work was provided by the National Research Council (CONICET) of Argentina, and by the ANPCYT of Argentina. O. C. and M. E. M. are members of the Scientific Research Career of the CONICET.

\newpage


\end{document}